\documentclass[twocolumn,showpacs,aps,prd,superscriptaddress,paperletter]
{revtex4}

\usepackage{graphicx}
\usepackage{dcolumn}
\usepackage{amsmath}
\usepackage{epsfig}
\usepackage{multirow}

\input babarsym

\def\mHH     {\ensuremath{m_{\pi\pi}}\xspace}
\newcommand{\DE} {$\mathrm{\Delta E}$}

\def\BpBm  {\ensuremath{\B^{\pm}}}
\def\PDG04{{\it Particle Data Group 2000}}

\def\jpsipipi  {\ensuremath{\jpsi \pi^+ \pi^-}\xspace}

\def\jpsipipiz  {\ensuremath{\jpsi \pi^+ \pi^0}\xspace}

\def\jpsirhoz {\ensuremath{\jpsi \rho^0}\xspace}
\def\jpsirhoch {\ensuremath{\jpsi \rho^+}\xspace}
\def\jpsiks {\ensuremath{\jpsi \KS}\xspace}
\def\jpsif {\ensuremath{\jpsi f_2}\xspace}

\newcommand{\BABARPubYear}    {07}
\newcommand{\BABARPubNumber}  {017}

\newcommand{\SLACPubNumber} {12441}

\begin{document}

\preprint{\babar-PUB-\BABARPubYear/\BABARPubNumber} 
\preprint{SLAC-PUB-\SLACPubNumber} 
\preprint{MAN/HEP/2007/5}

\begin{flushleft}
\babar-PUB-\BABARPubYear/\BABARPubNumber\\
SLAC-PUB-\SLACPubNumber\\
\end{flushleft}

\title{
{\large \bf
Branching fraction and charge asymmetry measurements in \boldmath{$B \ra \jpsi \pi \pi$} decays} 
}

%
\author{B.~Aubert}
\author{M.~Bona}
\author{D.~Boutigny}
\author{Y.~Karyotakis}
\author{J.~P.~Lees}
\author{V.~Poireau}
\author{X.~Prudent}
\author{V.~Tisserand}
\author{A.~Zghiche}
\affiliation{Laboratoire de Physique des Particules, IN2P3/CNRS et Universit\'e de Savoie, F-74941 Annecy-Le-Vieux, France }
\author{J.~Garra~Tico}
\author{E.~Grauges}
\affiliation{Universitat de Barcelona, Facultat de Fisica, Departament ECM, E-08028 Barcelona, Spain }
% ...
\collaboration{The \babar\ Collaboration}
\noaffiliation

\date{\today}%

\begin{abstract}

We study the decays $\Bz \ra \jpsipipi$ and $B^+ \ra \jpsipipiz$,
including intermediate resonances,
using a sample of 382 million \BB pairs recorded by the
\babar\ detector at the \pep2\ \epem $B$ factory. We measure the branching fractions
$\BR(B^0 \ra \jpsi \rho^0) = (2.7 \pm 0.3 \pm 0.17) \times 10^{-5} $ and
$\BR(B^+ \ra \jpsi \rho^+) = (5.0 \pm 0.7\pm 0.31) \times 10^{-5} $.

We also set the following upper limits at the 90\% confidence level: 
$\mathcal{B}(B^0 \ra \jpsi \pi^+ \pi^-$non-resonant) $< 1.2 \times 10^{-5}$,
$\mathcal{B}(B^0 \ra \jpsif) < 4.6 \times 10^{-6}$, and
$\mathcal{B}(B^+ \ra \jpsipipiz$ non-resonant)$ < 4.4\times
10^{-6}$.  We measure the charge
asymmetry in charged $B$ decays to $\jpsi \rho$ to be $-0.11 \pm
0.12 \pm 0.08$.

\end{abstract}


\maketitle

\begin{figure}[htb]
\begin{center}
\mbox{\epsfig{file=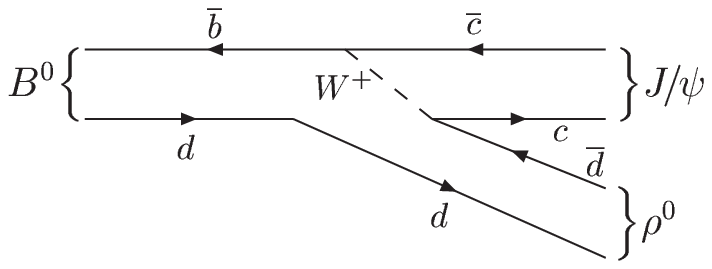,width=0.79\linewidth}}
\end{center}
\begin{center}
\mbox{\epsfig{file=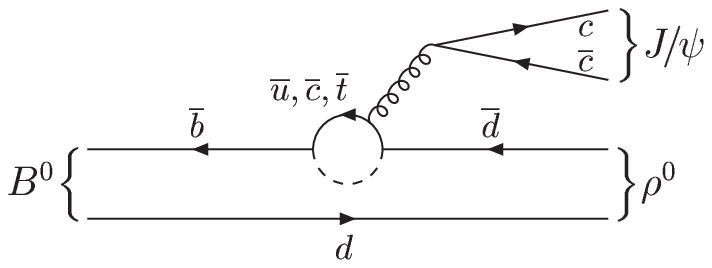,width=0.79\linewidth}}
\caption{Tree and penguin diagrams for the process $B^0 \ra \jpsi \rho^0$}
\label{fig:feyn}
\end{center}
\end{figure}

The decay $B^0 \ra \jpsi \rho^0$~\cite{chargeConj} can in principle be used to measure
the $CP$ violation parameter \stwob.  However, the measurement is not
as straightforward as for \jpsiks\ \cite{sin2bBabar,sin2bBelle}, because
it involves the decay of a pseudoscalar meson to two vector mesons,
resulting in both $CP$-odd and $CP$-even final states.  
Furthermore, the decay can proceed through either a 
color-suppressed tree diagram,
or a penguin diagram, both shown in Fig.~\ref{fig:feyn}, and
interference between them could result in
direct $CP$ violation \cite{dunietz}.  
Direct $CP$ violation may also occur in $B^+ \ra \jpsirhoch$
decays, where it would manifest itself as a non-zero
charge asymmetry:
\begin{equation}
\mathcal{A}_{CP} = \frac{N(B^- \ra \jpsi \rho^-) - N(B^+\ra \jpsi \rho^+)}
        {N(B^- \ra \jpsi \rho^-) + N(B^+\ra \jpsi \rho^+)}.
\label{eq:asym}
\end{equation}
The large intrinsic width of the $\rho$ meson necessitates an analysis
of a significant portion of the invariant mass spectrum of the dipion
system. 

The branching fraction for $\Bz \ra
\jpsipipi$  has previously been measured at \babar\ to be $(4.6\pm 0.7 \pm
0.6)\times 10^{-5}$ \cite{babarJpsipipi}, including a $\jpsi \rho^0$
component with a branching fraction of $(1.6 \pm 0.6 \pm 0.4)\times
10^{-5}$.  This measurement used a data sample containing
approximately 56 million \BB\ pairs, 
which is a subset of the sample used in this analysis.
The charged $B$ decay to $\jpsi \rho^+$ has not previously been
observed,  the CLEO collaboration set an upper limit 
$\mathcal{B}(B^+ \ra \jpsi \rho^+) < 7.7 \times 10^{-4}$ at the 90\% confidence
level \cite{CLEO}. 

The data sample used here contains 382 million \BB\ pairs collected 
with the \babar\ detector at the \pep2 
asymmetric-energy \epem storage ring, taken at a center-of-mass (CM)
energy equivalent to the mass of the 
\FourS\ resonance.  An additional data sample, corresponding to an
integrated luminosity of $36.8\invfb$, taken at a CM
energy 40\mev\ below the \FourS\ resonance, is used to study 
backgrounds from continuum $q\overline{q}$ production, where $q$ =
$u, d, s, c$. 

A detailed description of the \babar\ detector can be found 
elsewhere~\cite{NIM}. Charged-particle trajectories are measured by a 
five-layer silicon vertex tracker (SVT) and a 40-layer drift 
chamber (DCH) operating in a 1.5~T solenoidal magnetic
field.  A detector of 
internally reflected Cherenkov light (DIRC) is used for charged hadron
identification.  Surrounding this is a CsI(Tl) electromagnetic
calorimeter (EMC), and finally the instrumented flux
return (IFR) of the solenoid, which consists of layers of iron
interspersed with resistive plate chambers or limited streamer tubes. 

The \jpsi\ meson is reconstructed in decays to $l^+l^-$, where $l^{\pm}$
refers to a charged lepton, $e^{\pm}$ or $\mu^{\pm}$.
Electrons are selected on the basis of the ratio of 
EMC shower energy to track momentum, and the energy profile of the EMC
shower.  
For $\jpsi \ra \epem$, an attempt is made to recover energy losses
from bremsstrahlung, 
by looking for showers in the EMC close to those from the electron
candidates.  This procedure increases the selection efficiency for $\jpsi \ra
\epem$ candidates by approximately $30\%$ \cite{exclCharmon}.
The muon selection algorithm uses a neural network, for which the
most important input is the number of interaction lengths traversed in the IFR.
The lepton pairs are fitted to a common vertex and the invariant mass
of the combination is required to be in the range 2.98 (3.06) to
3.14\gevcc\ for the \epem\ ($\mu^+ \mu^-$) channels.
In order to reduce the background from $B^0 \ra \jpsi
K^{*0}(K^{*0}\ra K^+\pi^-)$ decays, 
charged pion candidates are required to satisfy stringent particle 
identification 
criteria, based on combined ionization energy 
loss (\dedx) in the DCH and SVT with the Cherenkov angle measured in
the DIRC.  

All tracks are required to originate close to the interaction point,
and to lie in polar angle ranges where particle identification
efficiency is well measured.  The allowed ranges
correspond to the geometric acceptances of the DIRC for pions, the
EMC for electrons, and the IFR for muons.

Neutral pion candidates are formed by combining pairs of
isolated showers in the EMC.  These are
required to spread over a minimum of three crystals, and to have an energy 
greater than $200\mev$.

To form a $B$ candidate, the reconstructed \jpsi\ is combined with 
either a pair of oppositely
charged pions, or a charged pion and a $\pi^0$, and a kinematic and
geometric fit is used to ensure that all final state particles are
consistent with coming from the same decay point.  In this fit, we
constrain the invariant mass of the $l^+l^-$ and the $\gamma \gamma$
to have the nominal mass of the \jpsi\ and $\pi^0$, respectively
~\cite{PDG06}.
The energy difference, \DE, between the candidate energy and the
single beam energy, $E_{beam}^{CM}$, (both in the CM frame) is expected to be
close to zero for signal events, and
is therefore required to be in the interval
$-40$ to $40\mev$ ($-60$ to $80\mev$) for $B^0$ ($B^+$) candidates,
corresponding to approximately $\pm 3\sigma$ of the \DE\ resolution.  
Note that the range is asymmetric for $B^+$ candidates because the 
$\pi^0$ in the final state gives rise to a tail on the low side of the
distribution, due to the EMC response to photons.  
For events where
more than one $B$ candidate passes the selection criteria, the
candidate with the smallest value of $|$\DE$|$ is chosen.

The branching fraction for each signal channel is obtained from:
\begin{equation}
\mathcal{B} = \frac{N_{sig}}{N_{\BB} \times \epsilon_{sig} \times
  \mathcal{B}(\jpsi \ra l^+ l^-)} ,
\label{eq:BF}
\end{equation}
where $N_{sig}$ and $\epsilon_{sig}$ are the observed yield and
selection efficiency, respectively, for a specific signal channel,
and $N_{\BB}$ is the number of $B$ meson pairs.
We assume that the
\FourS\ decays equally often into neutral and charged $B$ meson pairs.
The $\jpsi \ra l^+l^-$ branching fraction is taken to be $(11.87 \pm 0.12)\%$
\cite{PDG06}.  

We extract the signal yields for the \jpsirhoz, \jpsipipi\
non-resonant, and \jpsif\ channels by performing a fit on the
sample of reconstructed $B^0$ candidates.
We also perform a similar fit to the sample of charged $B$ candidates in
order to obtain the signal yields for the decay channels $B^+ \ra \jpsirhoch$ and
$B^+ \ra \jpsipipiz$ non-resonant.
The fits are two-dimensional, extended,
unbinned maximum likelihood fits to the distributions of \mes\ and \mHH,  Seven event categories
are considered: (i) $\jpsi \rho$ signal, (ii)
$\jpsi \pi \pi$ non-resonant signal, (iii) \jpsif\ signal, (iv) \jpsi \KS events,
(v) background events that do not contain a \jpsi\ (non-\jpsi
background), (vi) background events containing a \jpsi\ (inclusive
\jpsi\ background), and (vii) selected background channels that have
been studied in more detail (exclusive \jpsi\ backgrounds).  In the
fit to neutral $B$ candidates, the decay channels that comprise category (vii)
are $\jpsi K^{*0}$, $\jpsi K^{*+}$, $\jpsi
K_1(1270)$, $\jpsi K^+$, $\jpsi \rho^+$~\cite{jpsirhoch}, and $\jpsi
\pi^+$.
For the fit to charged $B$ candidates, the exclusive \jpsi\ background 
channels are $\jpsi K^{*0}$, 
$\jpsi K^{*+}$, $\jpsi K_1(1270)$, $\jpsi K^+$, $\jpsi \KS$, and
$\jpsi \KL$.
In both cases, these decay channels are not included in category (vi).
Of course, categories (iii) and (iv) are only present in the fit to
neutral $B$ candidates.  

A probability density function (PDF) is constructed for each category,
and the sum of these PDFs is used to fit the data.  The likelihood
function for
the total sample is the product of the PDF values for each candidate, multiplied by a
Poisson factor:
\begin{equation}
\mathcal{L} = \frac{1}{N!}e^{-N^{\prime}}(N^{\prime})^N
\prod_{i=1}^{N}\mathcal{P}_i ,
\end{equation}
where $N$ and $N^{\prime}$ are the numbers of observed and expected
events, respectively, and $\mathcal{P}_i$ is the value of the total PDF
for event $i$.   For all event categories except for the exclusive
\jpsi\ background, $P_i$ is a product of one-dimensional PDFs in \mes\
and \mHH.

\begin{figure}[htb]
\begin{center}
\mbox{\epsfig{file=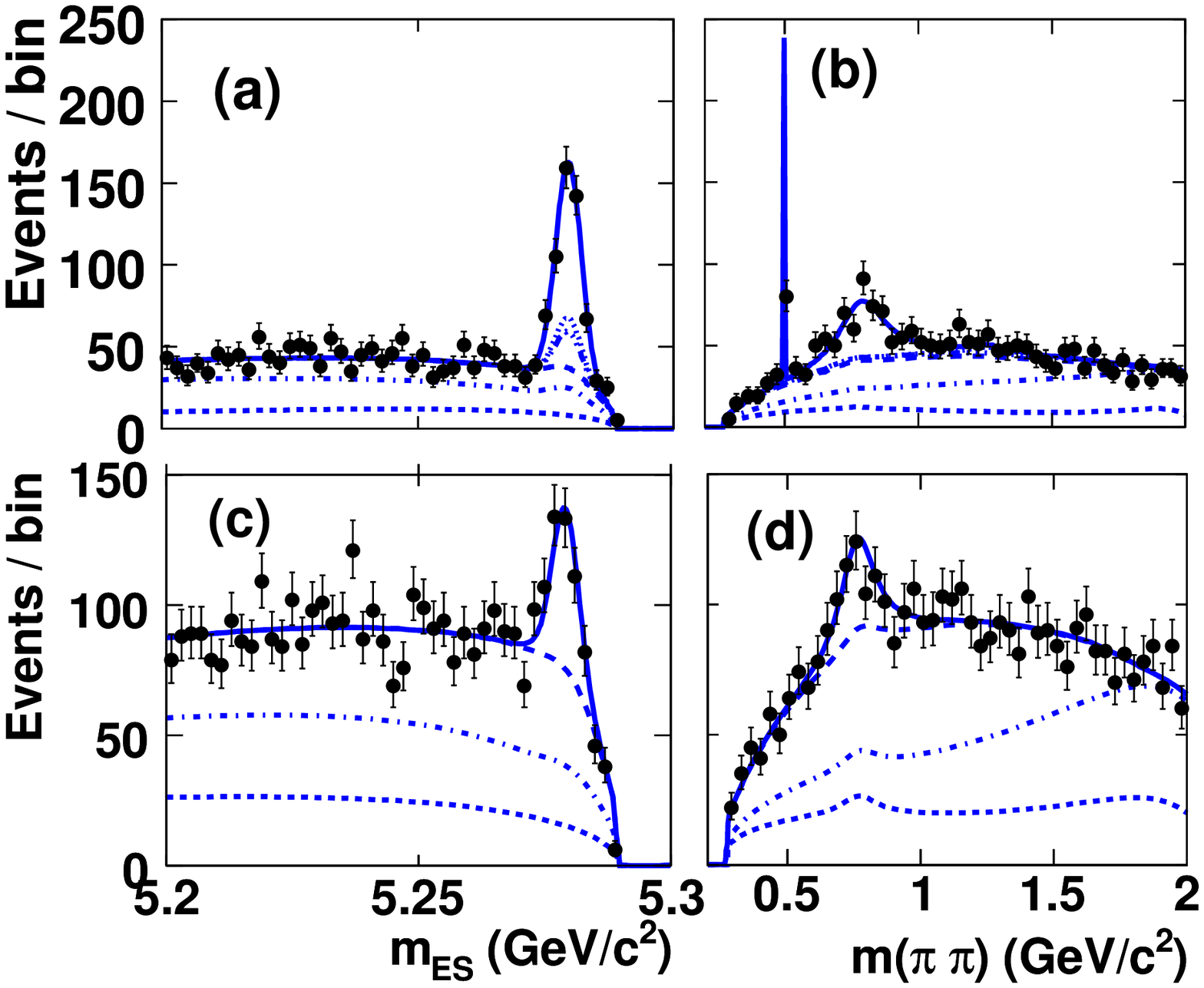,width=0.99\linewidth}}
\caption{Distributions of (a) \mes\ and (b) \mHH\ 
  for $B^0\ra \jpsipipi$ candidates.  The solid line
  represents the total PDF, while the other lines represent
  (cumulatively, from the bottom of the plot) non-\jpsi\ background,
  inclusive \jpsi\ background, exclusive \jpsi\ background, \jpsipipi\
  non-resonant signal, and \jpsif\ signal.  The points with error bars
  represent the data and statistical errors.  Plots (c) and (d) show the
  same distributions for $B^+ \ra \jpsipipiz$ candidates.  The sharp
  spike in (b) corresponds to \jpsiks\ events, while the broader peak
  is due to \jpsirhoz\ events.}
\label{fig:mesmpipi}
\end{center}
\end{figure}

Fig.~\ref{fig:mesmpipi} shows the \mes\ and \mHH\ distributions for
the data, and the projections of the PDFs for each category.  
The functional forms of these PDFs are as follows.
For the \jpsirhoz, \jpsipipi, \jpsif, and \jpsiks\ 
components, the \mes\ distributions are parametrized by Gaussian
functions, all with the same values for the mean and width, which are
allowed to float in the fit. 
In the fit to charged $B$ candidates, a Crystal Ball function \cite{CB}
is used instead for the \mes\ distributions of the
\jpsirhoch\ and \jpsipipiz\ signal components, as the presence of a
$\pi^0$ in the final state gives rise to a tail on the low mass side
of the peak. 

The \jpsi$\rho$ signal component is modeled by a relativistic $P$-wave
Breit-Wigner function~\cite{BW} in \mHH: 
\begin{equation}
F_\rho (\mHH)= \frac{\mHH
  \Gamma(\mHH)P^{2L_{eff}+1}}{((m_\rho^2-\mHH^2)^2+m_\rho^2 \Gamma(\mHH)^2)},
\label{eq:relBW}
\end{equation}
where $\Gamma(\mHH) =
\Gamma_0\left(\frac{q}{q_0}\right)^3\left(\frac{m_\rho}{\mHH}\right)\left(\frac
{1+R^2q_0^2}{1+R^2q^2}\right)$.  The parameter $q(\mHH)$ is the 
pion momentum in the
dipion rest frame, with $q_0 = q(m_\rho)$;  $P$ is the \jpsi\ momentum
in the $B$ rest frame;  $L_{eff}$ is the orbital angular momentum
between the \jpsi\ and the $\rho$ which can be 0, 1 or 2;  $R$ is the radius of the
Blatt-Weisskopf barrier factor~\cite{Blatt, radius}, which is taken to
be $(0.5 \pm 0.5)\fm$, and $m_{\rho}$ is the
$\rho$ meson mass.

The \mHH\ distribution for the $\jpsi \pi \pi$ non-resonant signal is 
$F_{\pi\pi} = q(\mHH)P^3$, the product of a
three-body phase space factor $q(\mHH)P$ and a factor
$P^2$ motivated by angular momentum conservation.

For the \jpsif\ component,  the \mHH\
distribution is described by a relativistic $D$-wave Breit-Wigner,
similar to Eq.~\ref{eq:relBW}, but with an extra factor
$(q/q_0)^2$ in the expression for $\Gamma(\mHH)$.

The decays to \jpsiks\ are not considered signal for this analysis.
Most of them are removed by the requirement that all tracks are 
consistent with coming from the
same vertex.
The \mHH\ distribution of the
remaining \jpsiks\ events are modeled by a narrow Gaussian function.

Non-\jpsi\ background events are modeled by an \mbox{ARGUS} function
\cite{argus} in \mes.  The \mHH\ PDF is the sum of two Weibull
functions \cite{wiebull}, and a Breit-Wigner to describe the 
$\rho$ component of the continuum background.  The parameters of this
PDF are fixed to values obtained from fits to the \jpsi\ mass sidebands of
the data.

The \mes\ distribution of the inclusive \jpsi\ background is an ARGUS
function plus a Gaussian at the $B$ mass.  The width of this Gaussian
is somewhat wider than that used for signal components as it
represents $B$ candidates that are not correctly reconstructed.
The \mHH\ PDF is a 4th-order polynomial.  The PDF parameters for this
component are fixed to values obtained by fits to a large sample of
$B\ra \jpsi(\ra l^+l^-) X$ Monte Carlo (MC) simulated events, with signal
events and exclusive \jpsi\ background channels removed.

Each of the exclusive \jpsi\ background channels is modeled 
by a two-dimensional PDF derived from the distribution of MC events
for that decay channel.  
The normalizations of these PDFs are 
determined by taking into account the selection efficiency on MC simulation,
and the world average branching fractions \cite{PDG06}.  

For the branching fraction fit to neutral $B$ candidates there are 
twelve free parameters: the yields of the $\jpsi \rho$, $\jpsi
\pi \pi$, $\jpsif$, \jpsiks, and inclusive \jpsi\ background
components, the mean and width of the Gaussian used for the signal
distribution in \mes, the parameters $m_{\rho}$, $\Gamma_0$, and
$L_{eff}$ in the $\rho$ lineshape, and the
mean and width of the \mHH\ distribution for the \jpsiks\ component.  
All other parameters are fixed, including those describing 
lineshape of the $f_2(1270)$, 
the normalizations and shapes of the 
exclusive \jpsi\ background PDFs, 
and the shapes of the inclusive \jpsi\ and non-\jpsi\ background PDFs.
We also fix the ratio of non-\jpsi\ to \jpsi\ (inclusive plus
exclusive) background yields
to a value obtained from fitting to data in the region $\mes < 5.26$
(i.e. lower in mass than the signal region), 
and extrapolated to the fit region using
distributions from MC simulation.

The configuration for the charged $B$ branching fraction fit is
very similar.  Here, there are eight free parameters, since there are no
\jpsiks\ or \jpsif\ components.

We find from MC simulation studies that correlations between
\mes\ and \mHH\ give rise to small biases in the
numbers of \jpsirhoch\ and \jpsipipiz\ non-resonant signal 
candidates found in the charged
$B$ fit.  The sizes of these biases are evaluated by examining the
distribution of residuals ($N_{obs} - N_{input}$) 
for a large number of MC experiments, and are listed in Table~\ref{tab:results}.
The yields obtained from the branching fraction fit are therefore corrected to
take account of this by subtracting these quantities from the fitted
yields.

The signal yields and statistical errors obtained from the branching
fraction fits 
are listed in Table~\ref{tab:results}.
We also list the statistical significances of the observed signals,
$\sqrt{-2\ln(\mathcal{L}_{Null}/\mathcal{L}_{Max})}$,
where $\mathcal{L}_{Max}$ is the likelihood from the fit, and 
$\mathcal{L}_{Null}$ is the value of the likelihood function
when the fit is performed with the signal yield constrained to zero 
events.

We obtain signal efficiencies using
samples of MC signal events, produced in monthly
blocks so as to 
match variations in detector and background conditions.  
Particle identification efficiency is
corrected using data control
samples of electrons, muons, and pions.  The sizes of
these corrections vary with momentum and polar angle, and average
corrections are
about $1.5\%$ for electrons, $5.9\%$ for muons, and $1.8\%$ for
pions.
With these corrections applied, about $85\%$ ($50\%$) of electron
(muon) pairs, and about $85\%$ of pions, satisfy their respective particle
identification requirements.
A small,
energy-dependent correction (typically about $-2\%$ relative) is also applied
to decay modes containing
a $\pi^0$ to account for known differences in photon detection
efficiency between data and MC simulation.  The 
corrected signal efficiencies are listed in Table~\ref{tab:results}.

\begin{table*}[!btp]
\caption{
Signal yields, detection efficiencies, and branching fractions
for the signal decay channels.  The fit bias, product of secondary
branching fractions ($\mathcal{B}(\jpsi \ra l^+l^-)$ and
$\mathcal{B}(\pi^0 \ra \gamma \gamma)$), and significances
of the signals (using statistical uncertainties only) are
also listed.  The corrected yields are obtained by subtracting the fit
bias from the fitted yields.  For the yields,
efficiencies, and branching fractions, the first errors are
statistical and the second are systematic.
For decay channels where no significant
signal is observed, we quote an upper limit at the $90\%$ confidence
level.}
\label{tab:results}
\begin{tabular}{lcccccc}
\hline \hline
Mode & Fit bias (events)& Corrected yield (events)& 
$\epsilon$(\%)&$\prod\mathcal{B}_i$(\%)&~Signif. ($\sigma$) &$\mathcal{B}
(\times 10^{-5})$ \\
\hline
\jpsirhoz & $0$ &$251.1 \pm 27.5 \pm 11.2$ & $20.6 \pm 0.1 \pm
0.8$& $\;\; 11.87
\pm 0.12$ & $13.0$ & $2.7 \pm 0.3 \pm 0.2$   \\ 
\jpsipipi & $0$ & $64.5 \pm 35.5 \pm 7.7$ & $20.3 \pm 0.1 \pm 0.8$
&$\;\; 11.87
\pm 0.12$ & $2.0$ & $<1.2$ (90\% C.L.) \\ 
\jpsif & $0$ & $24.4 \pm 13.8 \pm 1.8$ & $20.3 \pm 0.1 \pm 0.8$
&$\;\; 11.87
\pm 0.12$ & $2.0$ &  $<0.46$ (90\% C.L.)\\ \hline
\jpsirhoch & $-6.8 \pm 1.1$ & $218.5 \pm 28.8 \pm 9.5$ & $9.7 \pm 0.1 \pm 0.4$
&$\;\;11.73 \pm 0.12$ &$11.6$ &  $5.0 \pm 0.7 \pm 0.3$ \\ 
\jpsipipiz & $+8.2 \pm 0.9$ & $-12.7 \pm 27.1 \pm 4.7$ & $11.9 \pm 0.1 \pm 0.5$ &
$\;\;11.73 \pm 0.12$ & $0$ & $<0.44$ (90\% C.L.) \\ \hline \hline
\end{tabular}
\end{table*}

Systematic errors on the branching fraction measurements arise from
uncertainties on the signal efficiency, on the fitted yield, on
the number of \BzBzb\ or $B^+ B^-$  events in the sample, and on the $\jpsi \ra l^+
l^-$ branching fraction.  The number of \BB\ pairs is known to $1.1\%$
accuracy, and an additional $1.6\%$ uncertainty is assigned corresponding to the
assumption that the \FourS\ decays 50\% of the time into \BzBzb\ and
50\% of the time into $B^+ B^-$ \cite{PDG06}.  The fractional uncertainty
on $\mathcal{B}(\jpsi \ra l^+l^-)$ is $1.0\%$ \cite{PDG06}.

The systematic uncertainties on the efficiency are largely due
to imperfect simulation of the detector performance.  These
effects are studied using various data control samples.  
The largest sources of uncertainty are pion identification 
efficiency, a $2.0\%$ ($3.4\%$) relative error for charged (neutral) $B$ decay
channels, and $\pi^0$ efficiency ($3\%$ for
charged $B$ decays).  Tracking efficiency ($1.5\%$) and lepton
identification efficiency ($1.0\%$) also contribute to the uncertainty
on the efficiency.
The polarization of the $\rho$ in $B \ra \jpsi \rho$
decays is unknown. We use 
an MC sample in which the $\rho$ mesons are unpolarized to obtain the
central value of the signal efficiency. We also evaluate the
efficiency using
MC data samples with different $\rho$
polarizations, and observe a relative variation of 2\%, which is
assigned as a systematic uncertainty on the branching fraction
measurement.

We evaluate the impact of the fit procedure by observing the changes in
the yields when varying the PDF parameters that were fixed in the fit 
within their uncertainties.  The
resulting differences are added quadratically for sets of parameters
that are relatively uncorrelated,
and added linearly for highly correlated sets of parameters.  We also
repeat the fit using alternative functional forms for some PDFs, namely
the shape of the inclusive \jpsi\ background in \mHH, and the $\rho$
lineshape, and include the resulting differences in the yield in the
systematic uncertainty.  In addition, for the \jpsirhoch\ and
\jpsipipiz\ channels, systematic uncertainties equal to half of the 
bias corrections listed in Table~\ref{tab:results} are assigned.
The total systematic uncertainties on the yield
vary from 1.8 events for the \jpsif\ channel, to 11.2 events for the
\jpsirhoz\ channel.

In order to assess the charge asymmetry $\mathcal{A}^{\rho}$,  
we perform a second fit to the charged $B$ candidate sample.
In this fit, all the shape parameters for the signal
and background components are fixed to values obtained from the
branching fraction fit.  This reduces the number of free parameters
and improves the reliability of the fit.  
We include terms for the asymmetries in signal and background components
as follows:
\begin{eqnarray}
\mathcal{P}_i &=& N^{\rho} \times \frac{1}{2}(1-Q_i
  \mathcal{A}^{\rho})\mathcal{P}_i^{\rho} \nonumber \\
&+& N^{NR} \times \frac{1}{2}(1-Q_i
  \mathcal{A}^{NR})\mathcal{P}_i^{NR} \nonumber \\
&+& \sum_j N_j^{bkg} \times
  \frac{1}{2}(1-Q_i\mathcal{A}_j^{bkg})\mathcal{P}_{j,i}^{bkg} ,
\label{eq:asymPdf}
\end{eqnarray}
where $N^{\rho}$, $N^{NR}$, and $N_j^{bkg}$ are the yields for the
\jpsirhoch\ signal, the \jpsipipiz\ non-resonant signal, and the 
different background components $j$, respectively, $Q_i$ is the charge of
the  $B$ candidate in event $i$, and $\mathcal{A}^{\rho}$,
$\mathcal{A}^{NR}$, and $\mathcal{A}_j^{bkg}$ are the corresponding charge
asymmetries.
The asymmetry parameters for the exclusive \jpsi\ background channels are
fixed to world average values~\cite{PDG06}.  The asymmetries for the
non-\jpsi\ background and inclusive \jpsi\ background components are
assumed to be the same ($\mathcal{A}_{inc}^{bkg}=\mathcal{A}_{non}^{bkg} \equiv
\mathcal{A}^{bkg}$).  This fit therefore has
six free parameters:
the yields of the \jpsirhoch\ signal, \jpsipipiz\ non-resonant signal, 
and inclusive \jpsi\ background components, and the asymmetries 
$\mathcal{A}^{\rho}$,
$\mathcal{A}^{NR}$, and $\mathcal{A}^{bkg}$.

From the charge asymmetry fit, we obtain $A^{\rho} = -0.11
\pm 0.12 \mathrm{(stat.)}$.  The signal and background yields 
obtained from this fit are entirely consistent
with those from the branching fraction fit.

A potential contribution to the systematic uncertainty on the charge asymmetry
$\mathcal{A}^{\rho}$ could come from different pion identification
efficiencies for $\pi^+$ and $\pi^-$, leading to
different signal selection
efficiencies for positively and negatively charged $B$ candidates.
Using data control samples, this effect is found to be negligible.

The other sources of systematic error on the asymmetry are potential
differences in the backgrounds for positive and negative $B$
candidates.  The parameters describing the charge asymmetries 
of the exclusive \jpsi\ background
channels are varied within their uncertainties~\cite{PDG06}, assuming
a $10\%$ uncertainty for the $\jpsi K_1(1270)$ channel for which no
measurement is available.  The normalizations of the exclusive
background channels, and the shape parameters of the inclusive \jpsi
background and non-\jpsi\ background components are varied in turn, and
the fit is repeated.  The resulting changes to the
fitted value of $\mathcal{A}^{\rho}$ are added 
in quadrature, and the total systematic uncertainty is found to be
$\pm 0.08$.

In summary, we measure the following
branching fractions, where the first error in each case is statistical
and the second is systematic: $\mathcal{B}(B^0 \ra \jpsi \rho^0) = (2.7 \pm 0.3
\pm 0.2) \times 10^{-5}$, and $\mathcal{B}(B^+ \ra \jpsi \rho^+) = (5.0 \pm 0.7
\pm 0.3) \times 10^{-5}$.  The signals for $B^0
\ra \jpsif$, $B^0 \ra \jpsi \pi^+
\pi^-$ non-resonant, and $B^+ \ra \jpsi \pi^+\pi^0$ non-resonant are
not statistically significant, thus we set the following 
upper limits at the $90\%$ confidence level: 
$\mathcal{B}(B^0 \ra
\jpsif) < 4.6 \times 10^{-6}$, $\mathcal{B}(B^0 \ra
\jpsipipi) < 1.2 \times 10^{-5}$, and $\mathcal{B}(B^+ \ra
\jpsipipiz) < 4.4 \times 10^{-6}$.  These values are calculated by
summing the statistical and systematic uncertainties 
in quadrature, multiplying the result by 1.28, and adding it to the
central value of the branching fraction. 
We measure the charge asymmetry
defined in Eq.~\ref{eq:asym} for the decays $B^{\pm} \ra \jpsi
\rho^{\pm}$, $\mathcal{A}^{\rho} = -0.11 \pm 0.12 \pm 0.08$.

We are grateful for the excellent luminosity and machine conditions
provided by our \pep2\ colleagues, 
and for the substantial dedicated effort from
the computing organizations that support \babar.
The collaborating institutions wish to thank 
SLAC for its support and kind hospitality. 
This work is supported by
DOE
and NSF (USA),
NSERC (Canada),
IHEP (China),
CEA and
CNRS-IN2P3
(France),
BMBF and DFG
(Germany),
INFN (Italy),
FOM (The Netherlands),
NFR (Norway),
MIST (Russia),
MEC (Spain), and
PPARC (United Kingdom). 
Individuals have received support from the
Marie Curie EIF (European Union) and
the A.~P.~Sloan Foundation.


\begin{thebibliography}{99}
\bibitem{chargeConj} Charge conjugation is implied throughout this letter unless stated
otherwise.
\bibitem{jpsirhoch} Note that the $\jpsi \rho^+$ channel which is
  being measured in this note is one of the exclusive \jpsi\
  background channels in the fit to neutral $B$ candidates.  For the
  purpose of choosing the normalization for this PDF, we
  assume a value of $(6.0 \pm 6.0)\times 10^{-5}$ for the branching
  fraction.
\bibitem{wiebull}
The Weibull
function can be written as
\vspace{-0.3cm}
\begin{equation*}
W(m) = C\,V\,(m-M_{\rm on})^{(C-1)} \exp
[-V (m-M_{\rm max})^C] ,
\end{equation*}
where $V = (C-1)/(C(M_{\rm max}-M_{\rm on})^C)$.\\
$M_{\rm max}$ is the position of the function maximum, $M_{\rm on}$ is the
lower kinematic cut-off, and $C$ is a shape parameter.
\end{thebibliography}
\end{document}